\documentclass[aps,prl,reprint,floatfix,amsmath,amssymb,twocolumn]{revtex4-2}
 \usepackage{CJK}
\usepackage{epsfig}
\usepackage{graphicx}
\usepackage{dcolumn}
\usepackage{bm}
\usepackage[colorlinks=true,citecolor=blue, dvipdfm]{hyperref}
\usepackage{amssymb}
\usepackage{enumerate}

\begin{document}
\title{Fixed points and crossovers for the hysteresis scaling of dynamic mean-field models}

\author{Jiapeng Yang and Fan Zhong}
\email{stszf@mail.sysu.edu.cn}
\affiliation{School of Physics and State Key Laboratory of Optoelectronic Materials and
Technologies, Sun Yat-sen University, Guangzhou 510275, People's Republic of China}

\date{\today}

\begin{abstract}
Phase transitions are divided into first-order phase transitions  and continuous ones in current classification. While the latter shows striking phenomena of scaling and universality, the former is generically characterized by discontinuous jumps in extensive variables and pronounced hysteresis. Recent studies have demonstrated universal scaling behavior controlled by a cubic fixed point in first-order phase transitions. However, more recent investigations into the hysteresis in a dynamic mean-field quartic model driven through its first-order phase transitions have revealed new scaling exponents for different driving rates. Here, we discover a new exponent for large driving rates arising surprisingly from critical phenomena and show that, depending on the magnitude of the driving rates and on the absence or presence of noise, the same mean-field model remarkably exhibits several universality classes with definite universal scaling exponents governed by their corresponding fixed points through a systematic scaling analysis based on renormalization group theory. The theories and their various crossovers between different fixed points along with complete universal scaling of full curve collapse are verified by numerical results. This further confirms universal scaling in first-order phase transitions.
\end{abstract}

\maketitle

According to a modern classification~\cite{Fisher67}, phase transitions are divided into first-order phase transitions (FOPTs) and continuous ones. While it is well known that the latter is characterized by universality and scaling near their critical points~\cite{Mask,Cardyb,Justin,Amit}, whether FOPTs~\cite{Gunton83,Binder,Binder2,Binder16} also exhibit universal scaling behavior similar to their continuous counterparts is a long sought question~\cite{zhong18}. 

A characteristic of FOPTs is hysteresis, which usually occurs when an FOPT is driven with a finite rate through its equilibrium transition point. A typical example is the magnetization of a ferromagnet in which the magnetization changes from one orientation (phase) to the other when an externally applied field is varied. Power-law scaling of FOPTs and hysteresis with respect to the field parameters has long been reported both experimentally and theoretically~\cite{Rayleigh,Steinmetz,Rao,Rao1,Lo,Dhar,Jung,Rao1,Sengupta,Somoza,Mahato,Thomas,He,acha,Luse,zhongjpc,Jiang,zhongpret,zhongpre,Hohl,zhongprl,zhongssc,kim,Suen,chak,lee,Kuang,schulke,zhong02,Park,pan,Yildiz,zhu,zhongl05,Zuo,Cerruti,Fan,liyantao,zhong12,zhong16,Lee16,Lee16,liang,Pelissetto16,liange,zhong18,Bar,Pal,qiu,Kundu,zhong24,zhong24f,zhong25,zhangy,Wu,Chen,Chene}. In this paper we will focus on mean-field models only. For a noiseless mean-field quartic $\phi^4$ model, the dependence of the magnetic hysteresis on the field change rate can be exactly solved, giving rise to coercivity scaling with an exponent, referred to as hysteresis (scaling) exponent below, of $2/3$~\cite{Jung}, in agreement with numerical results~\cite{Luse,zhongjpc,zhongl05,zhong16} and that of field-like thermal transitions~\cite{liang}. However, when there is noise, numerical calculations several years ago found good power-law scalings over three orders of magnitude variation of the rate but with nonuniversal exponents depending on the noise amplitude~\cite{Kundu}, consistent with a previous finding that no single power law exists~\cite{Mahato}. Yet, devising a method of complete universal scaling whereby all parameters are  coordinately rescaled, it has been shown that all hysteresis curves themselves can be completely collapsed onto each other using universal exponents including the aforementioned $2/3$, indicating that the nonuniversal exponents are only effective when contributions from other parameters are neglected~\cite{zhong24,zhong24f}. Recently, it was found that apart from the $2/3$ exponent, the hysteresis also exhibits $1$ and $1/2$ exponents for small and large rates, respectively~\cite{Wu,Chen,Chene}. Here, we will show that these exponents, with $1/2$ replaced by a new one $3/5$ found herein arising unexpectedly from critical phenomena, are controlled by different fixed points with crossovers between them in the same single quartic mean-field model using a systematic scaling analysis based on renormalization-group theory. We also find that every fixed point in the mean-field theories can yield full curve collapse in all regimes controlled by different fixed points, including the crossovers between them, due to the absence of anomalous dimensions~\cite{Mask,Cardyb,Justin,Amit}. 

Consider a general model free energy
\begin{equation}
	f_n(\phi)=\frac{1}{2}a_2\phi^2+\frac{1}{n}a_n\phi^n-H\phi \label{f4}
\end{equation}
for an order parameter $\phi$ and its ordering field $H$, where $a_2$ is a reduced temperature and $a_n$ a general coupling constant. For $n=4$, the conventional quartic $\phi^4$ model exhibits a continuous transition at $a_2=0$ and $H=0$ but FOPTs for all $a_2<0$ between two ordered phases with $\phi_{\rm eq}=\pm\sqrt{-a_2/a_4}=\pm M_{\rm eq}$ at $H=0$. Dynamics can be studied by the usual Langevin equation~\cite{Hohenberg}
\begin{equation}
	\frac{\partial\phi}{\partial t}=-\frac{\partial f_n}{\partial\phi}+\zeta=-\left(a_2\phi+a_n\phi^{n-1}-H\right)+\zeta,\label{lang}
\end{equation}
where $\zeta$ is a Gaussian white noise whose averaged moments satisfy
\begin{equation}
\langle\zeta\rangle=0,\qquad  \langle\zeta(t)\zeta(t')\rangle=2\sigma\delta(t-t'),\label{zeta}
\end{equation}
with a noise amplitude $\sigma$ and we have set the conventional kinetic coefficient to unity as the time unit. Let the external field varies linearly with time as $H=H_{\rm in}+Rt$ with an initial field $H_{\rm in}$ and a constant rate $R$. Again for $n=4$, the transition from $-M_{\rm eq}$ to $M_{\rm eq}$ occur not at the equilibrium point $H=0$, but is delayed to a larger field $H>0$. This is hysteresis that increases with the rate $R$, since the system is farther out of equilibrium. Moreover, in a noiseless system, the transition can only take place beyond the so-called spinodal point ($H_s$, $M_s$), at which the free energy barrier between the two phases vanishes. As such, let~\cite{zhongl05,zhong16}
$\phi=M_s+\varphi$,
one finds $f_4(\phi)=f_4[M_s]+f_3(\varphi)+a_4\varphi^4/4$ with
\begin{equation}
f_3(\varphi)=\frac{1}{2}\tau\varphi^2+\frac{1}{3}a_3\varphi^3-h\varphi, \label{f3}
\end{equation}
where
\begin{equation}
	\tau=a_2+3a_4M_s^2,\quad h=H-H_s,\quad a_3=3a_4M_s,\label{tha3}
\end{equation}
the first two being an effective reduced temperature and an effective field, respectively, that become zero exactly at $M_s=\pm\sqrt{-a_2/3a_4}$, and $H_s=a_2M_s+a_4M_s^3=2a_2M_s/3$.
Therefore, $\tau=0$ and $h=0$ exactly at the spinodal point, similar to $a_2=0$ and $H=0$ at the critical point. This gives rise to a cubic $\phi^3$ theory near the spinodal where the quartic term is negligible. The dynamic equation becomes
\begin{equation}
	d\varphi/dt=-\left(\tau\varphi+a_3\varphi^{2}-h\right)+\zeta.\label{lang3}
\end{equation}
In the absence of the noise, Eq.~(\ref{lang3}) can be analytically solved by Airy's functions whose argument is proportional to $R^{-2/3}h$ for $h=Rt$. This gives rise to the exact mean-field $2/3$ hysteresis exponent~\cite{Jung}.

To see how this and other exponents emerge generally, we make a scale transformation to the dynamic equation, Eq.~(\ref{lang})~\cite{zhong24,zhongl05,Amit}. This is to change every quantity ${\cal O}$ to ${\cal O}'$ through ${\cal O}={\cal O}'b^{-[{\cal O}]}$ for a scaling factor $b$, where we have utilized the square brackets to denote the scale dimension. To keep the transformed dynamic equation identical with the original one, one must have
$[t]=-[a_2]$, $[H]=[\phi]+[a_2]$, $[\zeta]=[\phi]-[t]/2=[\sigma]/2$, and $[a_n]=[a_2]-(n-2)[\phi]$. According to the renormalization-group theory~\cite{Mask,Cardyb,Justin,Amit}, the Gaussian fixed point of a $\phi^n$ theory is reached by setting $[a_n]=0$. This yields
\begin{equation}
[\phi]=\frac{1}{n-2}[a_2],~[H]=\frac{n-1}{n-2}[a_2],~[\sigma]=\frac{n}{n-2}[a_2]\label{dim0}
\end{equation}
together with $[t]=-[a_2]$. The original and transformed equations must share the same solution since they are identical.
Accordingly,
\begin{equation}
\phi=b^{-[\phi]}\phi(a_2b^{[a_2]},a_nb^{[a_n]},Hb^{[H]},tb^{[t]},\sigma b^{[\sigma]}).\label{st}
\end{equation}
We then choose $H_{\rm in}=-Rt_0$ and set $t_0=0$ such that $H=R t$, replace $ t$ with $R$, and let $b=R^{-1/r}$ for $r\equiv[R]$, which satisfies~\cite{Gong}
\begin{equation}
r=[H]-[t]=(2n-3)[a_2]/(n-2).\label{rmf}
\end{equation}
These lead to the finite-time scaling (FTS) forms~\cite{Gong,Gong1,Huang},
\begin{equation}
	H=R^{\frac{n-1}{2n-3}}g_n(a_2R^{-\frac{n-2}{2n-3}},a_n,MR^{-\frac{1}{2n-3}},\sigma R^{-\frac{n}{2n-3}}),\label{nr}
\end{equation}
where $g_n$ (applicable to both a general $n$ and each specific value in the equations below) is a universal scaling function and $M=\langle\phi\rangle$ the order parameter averaged over the noise. Note that we have expressed $H$ as a function of $M$. The reason why Eq.~(\ref{nr}) is referred to as an FTS form is that $R^{[t]/r}=R^{-z/r}$ (derived from $tR^{-[t]/[r]}$ by substituting $b=R^{-1/r}$ into Eq.~(\ref{st})) is an externally imposed finite driving timescale, in direct analogy with the finite system size in finite-size scaling. This controllable finite timescale allows one to circumvent critical slowing down, just as the finite size allows one to overcome divergent correlation lengths~\cite{Gong,Gong1,Huang,Feng,Yuan}. 

Equation~(\ref{nr}) describes the general, systematic dependence of $H$ on $R$ and other parameters. This is the advantage of the renormalization-group analysis over direct approximations to the dynamic equation that usually just produce approximate relations to one specific parameter rather than the whole scaled variables. From Eq.~(\ref{nr}), we obtain the leading hysteresis exponent as $(n-1)/(2n-3)$, taking on values $1$, $2/3$, $3/5$, and $1/2$ for $n=2$, $3$, $4$, and $\infty$. It decreases from 1 to $1/2$ as $n$ increases from $2$ to infinity. Moreover, the corresponding FTS forms are
\begin{eqnarray}
	H&=&Rg_2(a_2,a_4R^{2},MR^{-1},\sigma R^{-2}),\label{phi2}\\
	H&=&R^{2/3}g_3(a_2 R^{-1/3},a_4R^{1/3},MR^{-1/3},\sigma R^{-1},a_3),\quad\label{phi3}\\
	H&=&R^{3/5}g_4(a_2R^{-2/5},a_4,MR^{-2/5},\sigma R^{-4/5}),\label{phi4}\\
	H&=&R^{1/2}g_{\infty}(a_2R^{-1/2},a_4R^{-1/2},M,\sigma R^{-1/2},a_{\infty})\label{phity}
\end{eqnarray}
by substituting the specific $n$ values. In Eqs.~(\ref{phi2})--(\ref{phity}), besides Eq.~(\ref{phi4}), we also keep the scaled variable associated with $a_4$ in all other equations since we will numerically solve Eq.~(\ref{lang}) for $n=4$. The corresponding exponent can be derived from $[a_4]=[a_3]-[\phi]$ for $n=3$ using Eq.~(\ref{tha3}) and $[a_4]=[a_2]-2[\phi]$ from the expression for general $n$. Indeed, for $n=2$, $[a_2]=0$ and so $[a_4]=-2[\phi]$, dictating the exponent relation between $a_4$ and $M$, while for $n=\infty$, $[\phi]=0$ and thus $[a_4]=[a_2]$, indicating that $a_4$ and $a_2$ share the same exponent. We have used $H$ and $M$ to express the scaling form for $n=3$ in Eq.~(\ref{phi3}). However, to observe the real cubic behavior, one must employ Eqs.~(\ref{f3}) and~(\ref{tha3}), resulting in
\begin{equation}
	h=R^{2/3}g_3'(a_2R^{-1/3},a_4R^{1/3},mR^{-1/3},\sigma R^{-1},a_3),\label{phi3r}
\end{equation}
where $g_3'$ is a scaling function different from $g_3$ and we have used directly $a_2$ instead of $\tau$ since they differ only by a constant of identical dimension~\cite{zhong24}. From Eqs.~(\ref{phi2})--(\ref{phi3r}), it is important to note that the exact hysteresis exponents can only be obtained if all the arguments in the scaling functions are kept constant; otherwise, effective exponents that depend on the parameters and $R$ result. In the following, we describe the mean-field behavior represented by each FTS form in Eqs.~(\ref{phi2})--(\ref{phi3r}), characterized by the marginal (dimensionless) $a_n$ and its corresponding Gaussian fixed point. This might suggest requiring a series of distinct models, each with its own highest order term $a_n$. Remarkably, however, we find that these FTS forms, except for the infinite-order case, collectively capture the distinct behaviors of the single quartic $\phi^4$ model itself across different $R$ and under both noisy and noiseless conditions. Moreover, they are all shown below to be full scaling forms because they can produce full curve collapse, or complete universal scaling.   

We now examine the FTS forms, Eqs.~(\ref{phi2})--(\ref{phity}), which will be referred to as the quadratic, cubic, quartic, and infinite-order behavior/scaling, respectively, for definite. First comes the quadratic behavior Eq.~(\ref{phi2}) in which the highest order term needed is $a_2$ corresponding to a quadratic $\phi^2$ model. The dynamic equation reduces to
\begin{equation}
	d\phi/dt=-a_2\phi+H+\zeta.\label{lang2}
\end{equation}
Indeed, one sees from Eq.~(\ref{phi2}) that $a_4$ is irrelevant and becomes negligible for $R\rightarrow0$. A normal situation this model describes is a paramagnet (the disordered phase) for $a_2>0$ or $T>T_c$, the Curie temperature. From Eq.~(\ref{lang2}), the equation of state is then simply $H=a_2M$ and the susceptibility is proportional to $1/a_2$, which is just Curie-Weiss' law. The hysteresis depends linearly on $R$, giving rise to an exact hysteresis exponent $1$, which can also be derived from the analytical solution of Eq.~(\ref{lang2}) for $H=Rt$. Another situation in which this model is applicable occurs when the noise amplitude $\sigma$ is large enough so that the system makes a transition to a regime in which it can freely jump from one potential well to the other for $a_2<0$~\cite{zhongpre}. As a result, the averaged magnetization is so small that the quadratic $\phi^2$ term is sufficient in this high-noise regime, leading again to the linear scaling found recently~\cite{Wu,Chen,Chene}. 

One might imagine yet another situation for such small driving rates that the transition is close to the equilibrium transition point at $H=0$ and $a_2<0$. If, in contrast to the above situation, the averaged magnetization could keep its mean-field value and so would not be small. We could then expand $M$ around $M_{\rm eq}$ at $H=0$, arriving at
\begin{equation}
	\frac{dm}{dt}=-(a_2+3a_4M_{\rm eq}^2)m-H+\zeta=2a_2m-H\label{phi2m}
\end{equation}
to the leading order. Note that the coefficient of the linear term in Eq.~(\ref{phi2m}) is finite, different from the cubic $\phi^3$ model at the spinodal point Eq.~(\ref{f3}), and twice as large as the coefficient of the original quadratic $\phi^2$ model, Eq.~(\ref{lang2}). In fact, these two quadratic models represent the lowest order behaviors of the ordered phase and the disordered phase, whose susceptibilities have a universal amplitude ratio of $2$ in the mean-field level~\cite{Amit,Cardyb,Justin,Mask}. The FTS scaling of this low-rate regime would change to
\begin{equation}
	H=Rg_2(a_2,a_4R^{2},(M-M_{\rm eq})R^{-1},\sigma R^{-2}),\label{phi21}\\
\end{equation}
Therefore, one could again have a $\phi^2$ theory with a linear scaling. However, we will see that in this regime, at least for the present mean-field model in which nucleation is not required, $M$ does not retain its mean-field value but rather becomes smaller for sufficiently small rates, leading to the high-noise regime.

Next we study the cubic behavior Eq.~(\ref{phi3}) for the cubic $\phi^3$ theory where $a_3$ is marginal. As constructed in Eq.~(\ref{f3}), this theory characterizes the behavior of the transition in the vicinity of the spinodal point ($H_s,M_s$) within the noiseless quartic $\phi^4$ model. For the corresponding noisy model, the transition can occur prior to reaching $H_s$ due to noise activation. Crucially, any finite noise can trigger the transition because the mean-field nature of the model requires no nucleation. One can still estimate a noise-shifted spinodal near which the cubic theory ought to be valid again. On the basis of this idea, it has been shown that all curves of different $R$ values can indeed collapse onto each other if all parameters of the original quartic model are consistently rescaled according to the FTS form, Eq.~(\ref{phi3})~\cite{zhong24}. This indicates that the observed noise-strength dependent hysteresis exponents~\cite{Kundu} are effective, because contributions from scaled variables other than $hR^{-2/3}$ in Eq.~(\ref{phi3}) or~(\ref{phi3r}) were not considered. Additionally, the $2/3$ hysteresis exponent has also been observed in the intermediate $R$ range by expanding the dynamic equation, Eq.~(\ref{lang}), around a reference point ($H_p, R_p$)~\cite{Chen,Chene}.

Now we consider the quartic behavior Eq.~(\ref{phi4}) for the quartic model. This is the standard mean-field model for critical phenomena near the critical point $a_2=0$ and $H=0$~\cite{zhongl05,zhong24}. However, at sufficiently large $R$ values, $M$ is large enough that the quadratic $a_2$ term can be neglected compared to the $a_4$ term, leading to effective criticality. The hysteresis of the FOPTs is therefore unexpectedly captured by the critical $\phi^4$ theory for continuous phase transitions, yielding the leading hysteresis exponent of $3/5$.

Finally comes the infinite-order behavior Eq.~(\ref{phity}) in which $a_{\infty}$ is marginal. For very large $R$, $M$ is large and hence the highest order term $a_{\infty}$ is dominant, which in practice may be approximated by using a very large $n$. The leading hysteresis exponent ought to be $1/2$. However, for the $\phi^4$ theory, the highest-order term is only $a_4$. As a result, the FTS of the hysteresis must be described by Eq.~(\ref{phi4}) and the leading hysteresis exponent must be $3/5$ instead.

We see that different $R$ ranges may be controlled by different fixed points. This indicates that crossovers between different regimes---controlled by their respective fixed points---are inevitable over a wide range of $R$ values. For a noiseless quartic $\phi^4$ model, two distinct regimes exist: one as $R\rightarrow0$, controlled by the cubic fixed point, and the other as $R\rightarrow\infty$, governed by the quartic fixed point, corresponding to small and large $R$ values, respectively. The behaviors within two regimes for various descriptions using different $n$ values are 
\begin{eqnarray}
	HR^{-1}&{\rm vs~}a_4R^2\rightarrow\left\{\begin{array}{lll}
		\left(a_4R^2\right)^{-\frac{1}{2}},H\rightarrow H_s, R\rightarrow0,\\
		\left(a_4R^2\right)^{-\frac{1}{5}}, H\rightarrow R^{\frac{3}{5}}, R\rightarrow\infty,\end{array}\right.\nonumber\\
		\label{phi2cr}\\
	HR^{-\frac{2}{3}}&{\rm vs~} a_2R^{-\frac{1}{3}}\rightarrow\left\{\begin{array}{lll}
		\left(a_2R^{-\frac{1}{3}}\right)^{2},H\rightarrow H_s, R\rightarrow0,\\
		\left(a_2R^{-\frac{1}{3}}\right)^{\frac{1}{5}}, H\rightarrow R^{\frac{3}{5}}, R\rightarrow\infty,\end{array}\right.\nonumber\\
		\label{phi3cr2}\\
	HR^{-\frac{2}{3}}&{\rm vs~} a_4R^{\frac{1}{3}}\rightarrow\left\{\begin{array}{lll}
		\left(a_4R^{\frac{1}{3}}\right)^{-2},H\rightarrow H_s, R\rightarrow0,\\
		\left(a_4R^{\frac{1}{3}}\right)^{-\frac{1}{5}}, H\rightarrow R^{\frac{3}{5}}, R\rightarrow\infty,\end{array}\right.\nonumber\\
		\label{phi3cr4}\\
	HR^{-\frac{3}{5}}&{\rm vs~} a_2R^{-\frac{2}{5}}\rightarrow\left\{\begin{array}{lll}
		\left(a_2R^{-\frac{2}{5}}\right)^{\frac{3}{2}},H\rightarrow H_s, R\rightarrow0,\\
		\left(a_2R^{-\frac{2}{5}}\right)^{0}, H\rightarrow R^{\frac{3}{5}}, R\rightarrow\infty,\end{array}\right.\nonumber\\
		\label{phi4cr}\\
	HR^{-\frac{1}{2}}&{\rm vs~} a_4R^{-\frac{1}{2}}\rightarrow\left\{\begin{array}{lll}
		\left(a_4R^{-\frac{1}{2}}\right)^{1},~~H\rightarrow H_s, R\rightarrow0,\\
		\left(a_4R^{-\frac{1}{2}}\right)^{-\frac{1}{5}}, H\rightarrow R^{\frac{3}{5}}, R\rightarrow\infty,\end{array}\right.\nonumber\\
		\label{phitycr}
\end{eqnarray}
where the leftmost exponents on the lefthand side indicate the specific $n$ values considered. We have also listed in Eq.~(\ref{phitycr}) the description employed $n=\infty$, Eq.~(\ref{phity}), and in Eqs.(\ref{phi3cr2}) and~(\ref{phi3cr4}) two different representations for $n=3$. In fact, they can be consolidated into
\begin{equation}
		HR^{-\frac{2}{3}}{~\rm vs~} \frac{a_2}{a_4}R^{-\frac{2}{3}}\rightarrow\left\{\begin{array}{lll}
			\left(\frac{a_2}{a_4}R^{-\frac{2}{3}}\right)^{1},~~H\rightarrow H_s, R\rightarrow0,\\
			\left(\frac{a_2}{a_4}R^{-\frac{2}{3}}\right)^{\frac{1}{10}}, H\rightarrow R^{\frac{3}{5}}, R\rightarrow\infty.\end{array}\right.\\		\label{phi3cr}
\end{equation}
In all the above representations, the powers of the parentheses are termed crossover (or more precisely, trans-regional, since crossover often refers to the intermediate region between the two regimes) exponents that are determined to yield the correct asymptotic behaviors. The $0$ crossover exponent in Eq.~(\ref{phi4cr}) indicates that the regime is described by the quartic fixed point itself. Note that for $R\rightarrow0$, $H\rightarrow H_s$, the spinodal field. 

To explicitly observe the cubic behavior, we have to work at the spinodal point ($H_s$, $M_s$) according to Eq.~(\ref{phi3r}). Therefore, one can find by interpolation, instead of the coercivity $H_c$ at $M=0$, the field $H_0$ at $M=M_s$, which approaches $H_s$ as $R\rightarrow0$. Plateaux ought to emerge for small $R$ following Eq.~(\ref{phi3r}), representing the cubic scaling. It then crossovers to the large $R$ $3/5$ behavior governed by the quartic fixed point similar to Eq.~(\ref{phi3cr4}). However, as the scaled arguments are now associated with $h$ and $m$ instead of $H$ and $M$, the crossover to the quartic behavior described by Eq.~(\ref{phi4}) is less neat compared to Eq.~(\ref{phi4cr}). Nevertheless, sufficiently far away from the crossover region, the behaviors in different regimes are expected to follow Eqs.~(\ref{phi2cr})--(\ref{phitycr}) for $H_0-H_s$ as well.

For a noisy $\phi^4$ model, in addition to the two regimes for the noiseless model, emerges a third regime---the high-noise regime---described by Eq.~(\ref{phi2}) controlled by the quadratic fixed point for $R\rightarrow0$. There are then new crossovers between this regime and the cubic or quartic regimes, which may emerge for small, intermediate, and large $R$, respectively, depending on the model parameters. These crossovers are given by
\begin{eqnarray}
	HR^{-1}&{\rm vs~}a_4R^2\rightarrow\left\{\begin{array}{lll}
		\left(a_4R^2\right)^{-\frac{1}{6}},~~H\rightarrow R^{\frac{2}{3}}, {\rm cubic},\\
		\left(a_4R^2\right)^{-\frac{1}{5}}, ~~H\rightarrow R^{\frac{3}{5}}, {\rm quartic},\end{array}\right.\nonumber\\
	\label{phi2cr34}\\
	HR^{-\frac{2}{3}}&{\rm vs~} \frac{a_2}{a_4}R^{-\frac{2}{3}}\rightarrow\left\{\begin{array}{lll}
		\left(\frac{a_2}{a_4}R^{-\frac{2}{3}}\right)^{-\frac{1}{2}}\!,H\rightarrow R, {\rm quadratic},\\
		\left(\frac{a_2}{a_4}R^{-\frac{2}{3}}\right)^{\frac{1}{10}}, H\rightarrow R^{\frac{3}{5}}, {\rm quartic},\end{array}\right.\nonumber\\
	\label{phi3cr24}\\
		HR^{-\frac{3}{5}}&{\rm vs~} a_2R^{-\frac{2}{5}}\rightarrow\left\{\begin{array}{lll}
		\left(a_2R^{-\frac{2}{5}}\right)^{-1}\!,H\rightarrow R, {\rm quadratic},\\
		\left(a_2R^{-\frac{2}{5}}\right)^{-\frac{1}{6}}\!, H\rightarrow R^{\frac{2}{3}}, {\rm cubic},\end{array}\right.\nonumber\\
	\label{phi4cr23}
\end{eqnarray} 
where we have used the consolidated form for the cubic behavior and the two crossovers to quartic behavior in the first two equations are the same as the identical situations in Eqs.~(\ref{phi2cr}) and~(\ref{phi3cr}).

Two remarks are in order. First, when the barrier between the considered two phases is large compared to the noise strength, the high-noise regime may require too small rates to be practically executable within reasonable computation time for solving Eq.~(\ref{lang}). In this case, the quadratic behavior is absent and hysteresis plateaux are observed~\cite{Chen,Chene}. The hysteresis behavior is then similar to the noiseless case, including the corresponding crossovers given by Eqs.~(\ref{phi2cr})--(\ref{phi3cr}).

Second, as mentioned above, the real cubic behavior can only be probed at the spinodal point, which, however, is not exactly solvable for a noisy system. As an estimation, we employ the Kramer rate formula~\cite{Kramer,Hanggi}
\begin{equation}
	{\cal K}=\frac{\omega_0\omega_{\rm b}}{2\pi}\exp\{-\Delta E_{\rm b}/\sigma\},\label{kramer}
\end{equation}
for an overdamped particle attempting to escape from a metastable potential well over a barrier $\Delta E_{\rm b}$ with angular frequencies $\omega_0$ and $\omega_{\rm b}$, respectively, given by the square root of the absolute value at the corresponding second derivative of the potential. The scale-dependent spinodal $H_s$ (we use the same symbol for both noiseless and noise systems) is then estimated with the criterion that the escaping time $1/{\cal K}=cR^{-z/r}$, viz. proportional to the driving time $R^{-z/r}$ with a ratio $c$~\cite{zhong18}. One may compute the magnetization at this estimated spinodal, fit its $R$-dependence, and thereby determine the corresponding $M_s$ following Eq.~(\ref{phi3r}). However, since Eq.~(\ref{phi3r}) involves other rate-dependent scaled variables that remain unfixed, such direct fits are rendered inaccurate, similar to usual procedures for determining hysteresis exponents. As a compromise, for a given $b$ and its derived $H_s$, we plot the cubic scaling of $(H-H_s)R^{-2/3}$ versus $a_2R^{-1/3}$ for all $H>H_s$ of a magnetization curve and directly search for a point where the slope matches the theoretical slope of a regime according to whether it is cubic, or quadratic or quartic, the latter two being given in Eq.~(\ref{phi3cr24}). The $M$ value at this point is identified as $M_s$, and the corresponding $H$ value is $H_0$~\cite{Yangjp}. Although this seems {\it ad hoc} and the estimated spinodal may not unique, its existence is consistent with the theory and thus confirming it. 

\begin{figure*}
\centerline{\includegraphics[width=\linewidth]{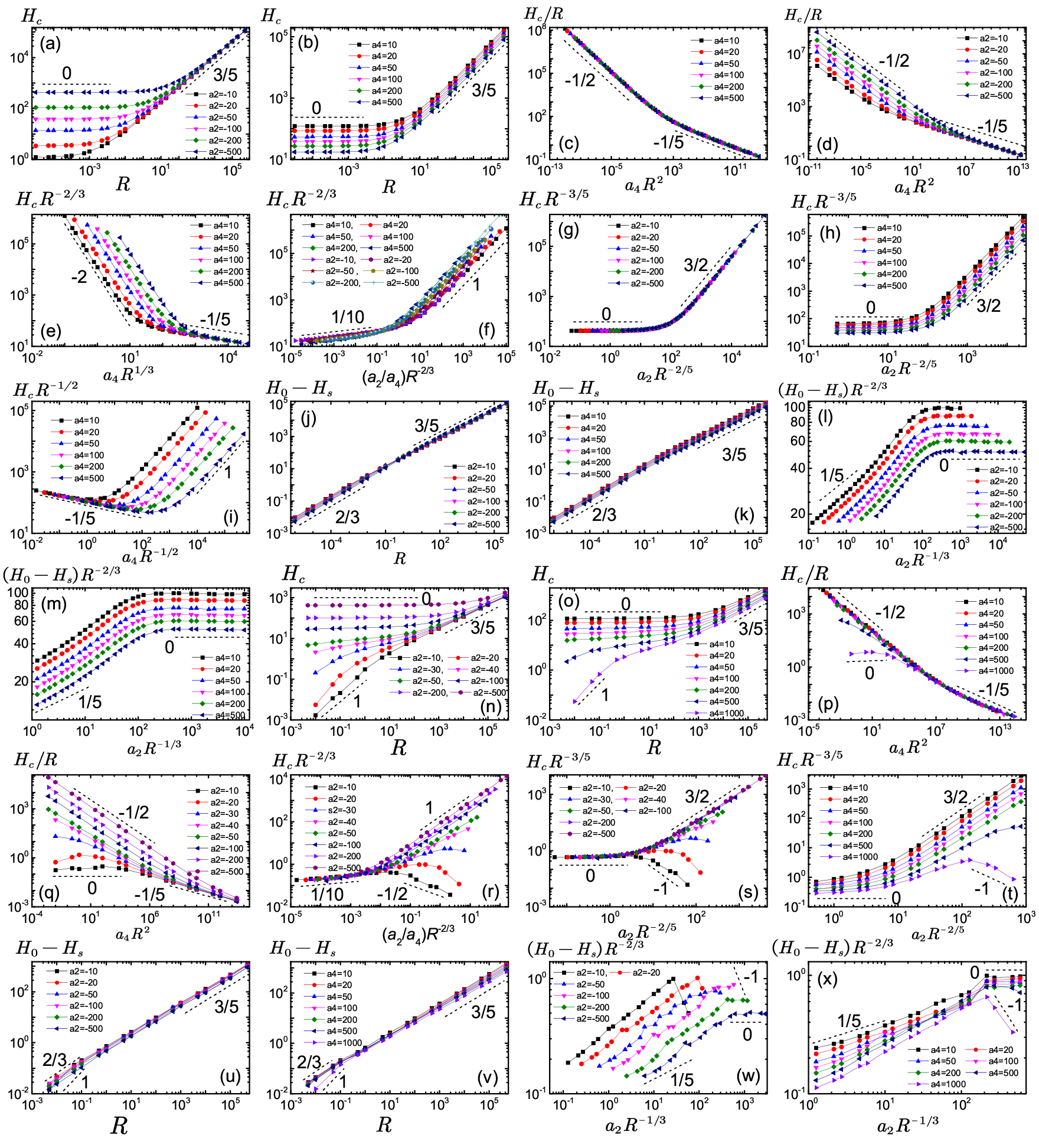}}
\caption{\label{fig1} (a) Coercivity $H_c$ versus $R$ at fixed $a_4=100$ with varying $a_2$. (b) $H_c$ versus $R$ at fixed $a_2=-100$ with varying $a_4$. Both (a) and (b) correspond to $\sigma=0$. Dashed lines with numerical labels in all panels indicate the theoretical (crossover) slopes from Eqs.~(\ref{phi2cr})--(\ref{phi3cr}) and~(\ref{phi3cr24}). (c) and (d) Quadratic rescaling of (a) and (b) by Eq.~(\ref{phi2}). (e) and (f) Cubic rescaling (a) and (b) by Eqs.~(\ref{phi3}) and~(\ref{phi3cr}). (g) and (h) Quartic rescaling of (a) and (b) by Eq.~(\ref{phi4}). (i) Infinite-order rescaling of (b) by Eq.~(\ref{phity}). (j) and (k) $H_0-H_s$ at $M_s$ versus $R$ for identical parameter sets as (a) and (b). (l) and (m) Cubic scaling of (j) and (k) by Eq.~(\ref{phi3r}). (n) and (o) $H_c$ versus $R$ for the same parameters as (a) and (b) but with $\sigma=0.4$. (p)--(t) Rescaling of (j) and (k) following Eqs.~(\ref{phi2})--(\ref{phi4}). (u) and (v) $H_0-H_s$ at the noisy $M_s$ for identical parameter sets as (n) and (o) using $c=50\pi$. (w) and (x) Cubic rescaling of (u) and (v) by Eq.~(\ref{phi3r}). Lines connecting symbols are only a guide to the eye.}
\end{figure*}

The noiseless dynamic equation, Eq.~(\ref{lang}), can be quite accurately solved using existing numerical recipe such as Runge-Kutta methods. For the noisy equation, we use explicit Euler's discretization with appropriate chosen time step to keep the solution stable. It is important to choose an initial field $H_{\rm in}$ sufficiently far away from the real transition point, the coercivity for example, to ensure independency of the initial conditions. 25000 samples were used for averaging. 

We show the dependence of $H_c$ on $R$ for the noiseless quartic model in Fig.~\ref{fig1}(a) and~\ref{fig1}(b) for fixed $a_4$ and $a_2$ but various $a_2$ and $a_4$, respectively. It is apparent that the slopes approach $0$ and $3/5$ for small $R$ and large $R$, respectively. This indicates that the system tends to constants for small $R$ and the behavior controlled by the quartic fixed point at large $R$. To confirm this, in Fig.~\ref{fig1}(c) and~\ref{fig1}(d) we depict the scaling plots following the quadratic behavior Eq.~(\ref{phi2}). One observes that $H_c$ for all $R$ and all $a_4$ collapses remarkably onto a single curve at fixed $a_2$, confirming that the quadratic scaling originates from fixed $a_2$. Meanwhile, the curves for different $a_2$ tend to distinct parallel lines at small $R$ but converge at large $R$---consistent with Eq.~(\ref{phi2}) since $g_2$ depends on $a_2$. Moreover, both small-$R$ and large-$R$ regimes are sloped lines (not horizontal), with corresponding slopes of $-1/2$ and $-1/5$. These values match the crossover exponents in Eq.~(\ref{phi2cr}), which governs crossovers to the constant $H_s$ at small $R$ and the $3/5$ scaling at large $R$. Similarly, scaling plots are shown for the cubic behavior in Figs.~\ref{fig1}(e) and~\ref{fig1}(f), the quartic behavior in Figs.~\ref{fig1}(g) and~\ref{fig1}(h), and the infinite-order behavior in~\ref{fig1}(i). Again, $H_c$ for all $R$ and all $a_2$ falls onto a single curve at fixed $a_4$, while curves of different $a_4$ remain parallel, as shown in Figs.~\ref{fig1}(g) and~\ref{fig1}(h). All curves exhibit two linear segments, whose slopes agree with the crossover exponents in Eqs.~(\ref{phi2cr})--(\ref{phitycr}). Horizontal lines appear only in Figs.~\ref{fig1}(g) and~\ref{fig1}(h), correctly reflecting the quartic behavior at large $R$. All other linear segments are oblique, representing crossovers to either the large-$R$ $3/5$ scaling or the small-$R$ constant $H_s$.

To explicitly reveal the leading $R$-dependence of the approach to $H_s$, we first find the field $H_0$ at $M_s$ instead of $H_c$ at $M=0$ and show its dependence on $R$ in Figs.~\ref{fig1}(j) and~\ref{fig1}(k). One sees two linear segments with slopes of $2/3$ and $3/5$ for small and large rates, respectively. Indeed, through the cubic rescaling of $H_0$ by Eq.~(\ref{phi3r}) in Figs.~\ref{fig1}(l) and~\ref{fig1}(m), the plateaux at small $R$, as expected, emerges consistent with the cubic behavior with the $2/3$ hysteresis exponent. The large $R$ portion exhibits a slope of about $1/5$ in agreement with the crossover to the quartic behavior, Eq.~(\ref{phi3cr2}). However, the agreement and linearity for large $R$ are not as good as Fig.~\ref{fig1}(f) (note the half slope due to the consolidated form) because of the effects of $H_s$ and $M_s$.

We have shown both from the theory and from the numerical results that the large $R$ behavior is the quartic $3/5$ critical scaling, different from the $1/2$ behavior found in Refs.~\cite{Chen,Chene}. The infinite-order rescaling in Fig.~\ref{fig1}(i) confirms the absence of such a behavior through the absence of a plateau. It is probably an effective exponent arisen from the initial conditions, which lead to additional scaled variables in the scaling functions if not in quasi-equilibrium, as analyzed in Ref.~\cite{Feng}. The $H_{\rm in}$ was chosen to be near the symmetrically opposite $H_s$~\cite{Chennote}, which is not sufficient for large rates as $M$ will rise directly rather than stabilize for some time in the metastable state following a short transient. Another evidence is that $1/2$ falls within $0$ and $3/5$, the range of exponents in Figs.~\ref{fig1}(a) and~\ref{fig1}(b). 

When the noise strength is finite, the dependence of $H_c$ on $R$ is distinct, as shown in Fig.~\ref{fig1}(n) and~\ref{fig1}(o) for a series of $a_2$ and $a_4$, respectively, with fixed $\sigma$. One sees that, besides the large-$R$ $3/5$ scaling and the small-$R$ plateaux, for $a_2$ so large or $a_4$ so small that the equilibrium barriers between the two phases $a_2^2/2a_4$ are small, the small $R$ portions exhibit linear decreases. This linearity at $M=0$ rather than $M_{\rm eq}$ rules out Eq.~(\ref{phi21}) as its origin. In addition, small rather than equilibrium $M$ is evident in Ref.~\cite{Chene}. We present $H_c$ in terms of the quadratic behavior in Figs.~\ref{fig1}(p) and~\ref{fig1}(q), cubic behavior in Fig.~\ref{fig1}(r), and quartic behavior in Figs.~\ref{fig1}(s) and~\ref{fig1}(t). The main curves in all these figures are similar to the corresponding noiseless behaviors in Figs.~\ref{fig1}(c)--\ref{fig1}(h) (missing~\ref{fig1}(e)), except that some curves in these figures progressively deviate from the main curves as the barriers between the two phases are lowered. The curves exhibiting large deviations show plateaux at small $R$ in the quadratic rescaling, thus confirming the linear scaling. At large $R$, they converge with the main curves demonstrating lines with slopes of $-1/5$, $1/10$, and $0$ in quadratic, cubic, and quartic rescaling, in accordance with the $3/5$ scaling behavior by Eqs.~(\ref{phi2cr}),~(\ref{phi3cr}), and~(\ref{phi4cr}), respectively, similar to the noiseless case. At small $R$, the main curves show sloped lines with slopes of $-1/2$, $1$, and $3/2$ for the quadratic, cubic, and quartic rescaling, respectively, according to the same equations, again identical to the noiseless cases. This agrees with the plateaux of the main curves at small $R$ in Figs.~\ref{fig1}(n) and~\ref{fig1}(o).

The identical features exhibited by the main curves in Figs.~\ref{fig1}(p)--\ref{fig1}(t) (excluding curves showing the additional quadratic scaling) as compared to Figs.~\ref{fig1}(c)--\ref{fig1}(h) for the noisy and noiseless behaviors unequivocally demonstrate that the plateaux in Figs.~\ref{fig1}(n) and~\ref{fig1}(o) must originate from the cubic fixed point for the spinodal. We therefore employ Eq.~(\ref{kramer}) and the criterion of comparable escape time and the driving time to estimate the noisy $H_s$ and $M_s$~\cite{Yangjp}. The resultant $H_0-H_s$ vs $R$ is depicted in Figs.~\ref{fig1}(u) and~\ref{fig1}(v). Two distinct slopes of $3/5$ and $2/3$ at the large and small $R$ regimes are evident. Even the linear $1$ slope is discernible at small $R$ for the quadratic scaling of large $a_2$ and $a_4$. Cubic rescaling of $H_0-H_s$ is displayed in Figs.~\ref{fig1}(w) and~\ref{fig1}(x). Plateaux for small $R$ emerge and transition to the quartic $3/5$ regime with a slope of $1/5$ given by Eq.~(\ref{phi3cr4}) or~(\ref{phi3cr24}). Crossing over from the cubic to quadratic scaling of the slope $-1$, Eq.~(\ref{phi3cr24}), is visible for $a_4=500$ in Fig.~\ref{fig1}(x) and ought to be evident from the progressive deviation from the main curves in Figs.~\ref{fig1}(n) and~\ref{fig1}(o). Direct crossover from the quartic to quadratic scaling with respective slopes of $-1/5$ and $-1$ as per Eq.~(\ref{phi3cr24}) and thus absence of the cubic behavior is also manifest in Figs.~\ref{fig1}(w) and~\ref{fig1}(x). However, this crossover appears abrupt since the data points can be chosen with some freedom, even though it is unequivocal in Figs.~\ref{fig1}(n)--\ref{fig1}(t). We simply display several representative data points as an illustration of the capacity of the current method to even detect the presence of the quadratic regime.

We have seen that all the scaling forms, Eqs.~(\ref{phi2})--(\ref{phi3r}), describe well all their respective regimes and the crossovers. One can then fix all scaled arguments in the scaling functions except for one and collapse all curves of different $R$ values. This has been performed in Ref.~\cite{zhong24} and confirmed in Ref.~\cite{Chene}. All the whole magnetization curves of all $R$ values---both large and small---collapse onto each other using Eq.~(\ref{phi3}) for both the noiseless and noisy quartic model and thus was termed complete universal scaling~\cite{zhong24}. However, as we have shown that the large and small $R$ regimes across an extended $R$ range may also obey distinct scaling behaviors, we therefore further investigate this complete universal scaling here.

\begin{figure}
\centerline{\includegraphics[width=0.8\linewidth]{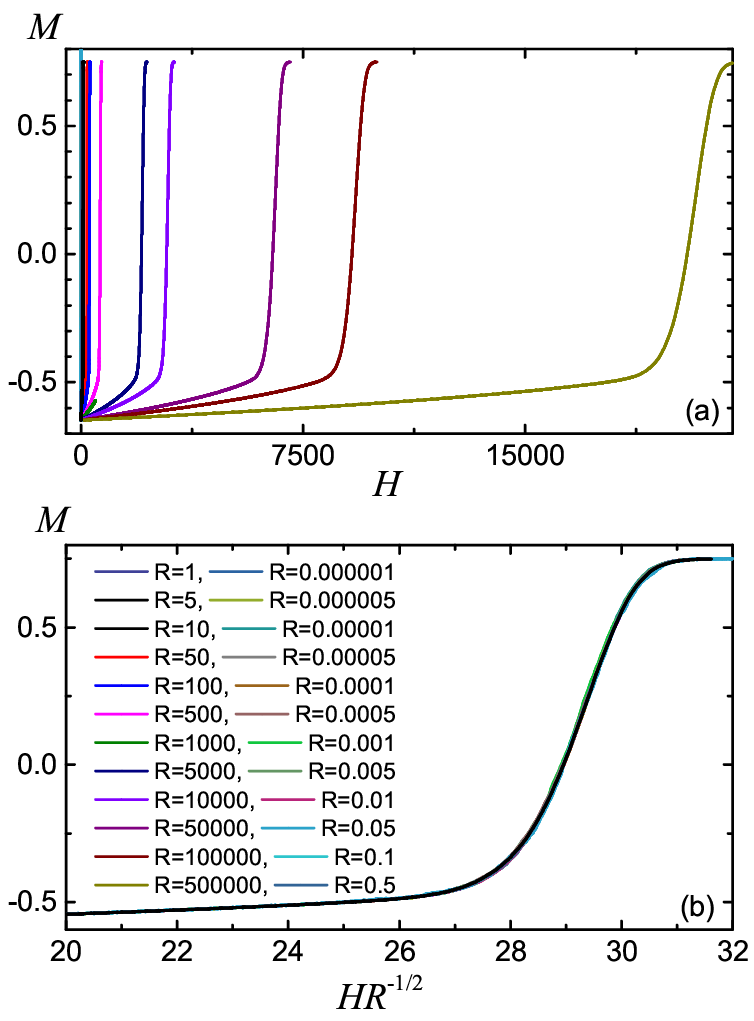}}	\caption{\label{fig2} (a) Magnetization curves for a series of $R$ values, as listed in the legend of (b). The curves shift to the right with increasing $R$. $a_2=a_{20}(R/R_0)^{-1/2}$, $a_4=a_{40}(R/R_0)^{-1/2}$, and $\sigma=\sigma_0(R/R_0)^{-1/2}$, $a_{20}=-400$, $a_{40}=948$, $\sigma_0=0.4$, $R_0=10$. (b) Rescaled curves from (a) based on the infinite-order behavior, Eq.~(\ref{phity})}.
\end{figure}
As an example, we show in Fig.~\ref{fig2}(a) the magnetization curves for a series of $R$ values varying over $11$ plus orders of magnitude with fixed $a_2R^{-1/2}$, $a_4R^{-1/2}$, and $\sigma R^{-1/2}$. The range is so large that the curves of small rates appear crowded together. After being rescaled according to the infinite-order behavior, Eq.~(\ref{phity}), all curves in Fig.~\ref{fig2}(a) completely collapse onto each other as seen in Fig.~\ref{fig2}(b). Note that the curves are solutions of the quartic model, Eq.~(\ref{lang}), rather than an infinite-order model. However, once the curves are rescaled with the scaled variables fixed according to the infinite-order behavior, Eq.~(\ref{phity}), they completely coincide, regardless of the regimes or the crossovers between them, and whether the regime is dominated by the quadratic, the cubic, or the quartic fixed point. We have checked that similar result applies to all other scaling forms for both noiseless or noisy systems. Therefore, any scaling form in Eqs.~(\ref{phi2})--(\ref{phity}) can completely collapse all curves in all regimes and their crossovers. This is reasonable as all the scaling forms keep the dynamic equation unchanged. It is thus not possible to distinguish different regimes by complete universal scaling. However, this does not mean that the method is useless. It is a full use of the scaling forms and thus completely verifies them. Moreover, we have checked that once spatial extensions are taken into account, the anomalous dimensions lift this over-universal behavior of the mean-field theory. It can then effectively avoid approximate estimation of the noisy $H_s$ and $M_s$, as demonstrated in Refs.~\cite{zhong24,zhong24f,zhong25}. 

We have systematically studied the dynamic quartic mean-field model driven by a linearly varying external field across its first-order phase transitions and shown that the large rate scaling exponent is $3/5$ governed strikingly by the quartic fixed point for critical phenomena instead of the recently found $1/2$. Employing the scaling analysis based on renormalization-group theory, we have developed systematic scaling theories for the diverse universal scaling behaviors remarkably exhibited by the mean-field model and controlled by distinct fixed points for various driving rates and noise conditions (absent/present). In the absence of the noise, the cubic and the quartic fixed points controlled the small and the large scaling behaviors with the leading hysteresis exponents of $2/3$ and $3/5$, respectively. On the other hand, in the presence of the noise, the additional quadratic fixed point with the leading exponent of $1$ may emerge if the driving rate is sufficiently low, while the cubic fixed point may be absent for a sufficiently strong noise. Crossover features between various regimes governed by the fixed points have also been analytically described and numerically validated within the theories. The spinodal points are important for the cubic behavior to emerge and have been estimated in the noisy model. We have also demonstrated that every scaling form of the fixed point can yield the complete universal scaling of full curve collapse in all different regimes and their crossovers due to the absence of anomalous dimensions in the mean-field theories, further corroborating the scaling theories and hence universal scaling behavior in first-order phase transitions.

\begin{acknowledgments}
We thanked Dr. Y.-H. Ma for informing us their initial conditions in their solutions. This work was supported by National Natural Science Foundation of China (Grant No. 12175316).
\end{acknowledgments}

\end{document}